# Electron Kinetics in a Positive Column of AC Discharges in a Dynamic Regime

Nathan A Humphrey[1] and Vladimir I Kolobov[2]

[1]*The University of Alabama, Tuscaloosa, AL 35487*

[2]*The University of Alabama in Huntsville,* Huntsville, AL 35899

## Abstract

We have performed hybrid kinetic-fluid simulations of a positive column in AC Argon discharges over a range of driving frequencies *f* and gas pressure *p* for the conditions when the spatial nonlocality of the Electron Energy Distribution Function (EEDF) is substantial. Our simulations confirmed that the most efficient conditions of plasma maintenance are observed in the dynamic regime when time modulations of mean electron energy (temperature) are substantial. The minimal values of the root mean square (rms) electric field and the electron temperature have been observed at *f/p* values of about 3 kHz/Torr in a tube of radius $R = 1$ cm. The ionization rate and plasma density reached maximal values under these conditions.

The numerical solution of a kinetic equation allowed accounting for the kinetic effects associated with spatial and temporal nonlocality of the EEDF. Using the *kinetic* energy of electrons as an independent variable, we solved an anisotropic tensor diffusion equation in phase space. We clarified the role of different flux components during electron diffusion in phase space over surfaces of constant *total* energy. We have shown that the kinetic theory uncovers a more exciting and rich physics than the classical ambipolar diffusion (Schottky) model. Non-monotonic radial distributions of excitation rates, metastable densities, and plasma density have been observed in our simulations at $pR > 6$ Torr cm. The predicted off-axis plasma density peak in the dynamic regime has never been observed in experiments so far. We hope our results stimulate further experimental studies of the AC positive column. The kinetic analysis could help uncover new physics even for such a well-known plasma object as a positive column in noble gases.

## I. Introduction

A positive column is a classical object of gas discharge physics with essential applications in lighting and gas lasers [1]. The low-temperature plasma of the positive column can be axially uniform or striated depending on the gas type and discharge conditions. The plasma can be maintained by direct current (DC) or alternating current (AC) power sources over various driving frequencies and gas pressures. The plasma properties are defined by the balance of electron impact ionization and the ambipolar diffusion to the wall or volume recombination. Two distinct time scales determine the dynamic properties of positive column plasma: the ambipolar diffusion time, $\tau_a$ and the electron energy relaxation time $\tau_u$. Depending on the AC angular frequency $\omega$, three operating regimes have been identified [2]. In the low-frequency regime, $\omega\tau_a < 1$, the axial electric field, $E_z$, maintaining the plasma remains nearly constant during most of the AC period, and the plasma density follows the current oscillations (quasi-DC case). In the high-frequency regime, $\omega\tau_u > 1$, plasma density remains constant, and the electric field follows the current oscillations with a phase shift that depends on the ratio $\nu/\omega$ of the electron collision frequency $\nu$ to the driving frequency. The intermediate case $\tau_a^{-1} < \omega < \tau_u^{-1}$ corresponds to a "dynamic regime," which is a focus of the present paper.

The dynamic regime found practical application in light sources: a noticeable increase of light intensity was observed in this regime [1, 3]. Experimental studies of the positive column of AC discharges in Helium revealed surprising off-axis peaks of metastable atom density [4], which were attributed to peculiarities of electron kinetics and ionization processes in pulsed plasma. The off-axis peaks of excitation rates were first observed in simulations of DC positive columns [5, 6] and later explained theoretically [7] by peculiarities of nonlocal electron kinetics in noble gases at elevated gas pressures. The dynamic regime of an inductively coupled plasma (ICP) has been studied in [8] using a fluid model for electrons.

The present paper aims to study electron kinetics and ionization processes in the dynamic regime under conditions where the spatial nonlocality of electron kinetics may be significant. Kinetic simulations of nonlocal electron kinetics in a steady-state DC positive column, where the energy and radial dependence of the Electron Energy Distribution Function (EEDF) cannot be factorized $f_0(u,r) \neq f_0^0(u) n_e(r)$, have been previously reported (see Refs. [9,10] and references therein). Many publications have been devoted to solving the *local* Boltzmann equation for the EEDF and analysis of temporal nonlocality when the EEDF is a *nonlocal* function of time $f_0(u,t) \neq f_0^0(u) n_e(t)$, [1]. However, no analysis of spatially nonlocal electron kinetics under conditions of temporal nonlocality, where the EEDF is a complex function of all three arguments, has been performed so far. The present paper aims to address this deficiency for an AC positive column in a collision-dominated regime when the electron mean free path is smaller than the tube radius. We discuss the peculiarities of electron kinetics and ionization processes in this system and confirm that the optimal conditions for plasma maintenance occur in the dynamic regime.

## II. Computational model

The hybrid model used in the present paper was first described in [11] and later adapted in [12] to study plasma stratification in noble gases. Here, this model is used to study the AC positive column in the collisional plasma of noble gases.

### 1. Kinetic Equation for Electrons

By using the two-term spherical harmonics expansion for the electron distribution function, $f_e = f_0 + \frac{v}{v} \cdot \boldsymbol{f}_1$, the Boltzmann kinetic equation for electrons in weakly ionized collisional plasma can be reduced to a set of coupled differential equations describing the isotropic component $f_0$ and anisotropic component $\boldsymbol{f}_1$ (sometimes called Davydov–Allis system) [13,14,15]:

$$\frac{\partial f_0}{\partial t} + \frac{v}{3} \nabla \cdot \boldsymbol{f}_1 - \frac{1}{3v^2} \frac{\partial}{\partial v}\left(v^2 \frac{e\boldsymbol{E}}{m} \cdot \boldsymbol{f}_1\right) = S_0 \qquad (1)$$

$$\frac{\partial \boldsymbol{f}_1}{\partial t} + v \nabla f_0 - \frac{e\boldsymbol{E}}{m} \frac{\partial f_0}{\partial v} = -\nu(v)\boldsymbol{f}_1 \qquad (2)$$

Here, $\nabla$ denote the divergence and gradient operators in configuration (physical) space, $e$ and $m$ are the unsigned electron charge and electron mass, $\mathbf{E}$ is the electric field vector, $\nu(v)$ is the transport collision frequency, and $S_0$ is the collision term describing energy loss in collisions and generating new electrons.

Using the electron kinetic energy $u = mv^2/(2e)$ expressed in eV, Eq. (1) can be rewritten in the form

$$\frac{\partial f_0}{\partial t} + \nabla \cdot \boldsymbol{\Phi} - \frac{1}{v} \frac{\partial}{\partial u}(v\Gamma) = S_0, \qquad (3)$$

where

$$\boldsymbol{\Phi} = \frac{v}{3} \boldsymbol{f}_1 \quad , \qquad \Gamma = \boldsymbol{E} \cdot \boldsymbol{\Phi} \qquad (4)$$

is the spatial flux, and $\Gamma$ is the flux along energy in phase space. We consider conditions when the angular frequency of the electric field variation $\omega$ is low compared to $\nu(v)$. In this case, the spatial flux becomes

$$\boldsymbol{\Phi} = -D_r \left(\nabla f_0 - \boldsymbol{E} \frac{\partial f_0}{\partial u}\right) \qquad (5)$$

where $D_r = v^2/(3\nu)$ is the spatial diffusion coefficient. Substituting (4) into (3), we obtain the kinetic equation for $f_0$ in the form of a diffusion equation in the 4-dimensional phase space ($r$, $u$):

$$\frac{\partial f_0}{\partial t} - \nabla_4 \cdot (\mathbf{D}\nabla_4 f_0) = S_0 \qquad (6)$$

Here $\nabla_4$ denotes the divergence and gradient operators in a 4-dimensional phase space ($\boldsymbol{r}, u$). The speed $v$ becomes the Lame coefficient, and $\mathbf{D}$ is the tensor:

$$\mathbf{D} = D_r \begin{pmatrix} I_3 & -\boldsymbol{E} \\ -\boldsymbol{E} & E^2 \end{pmatrix} \qquad (7)$$

where $I_3$ denotes the unit tensor in configuration space. The kinetic equation (6) has been used (in 2D phase space) for the analysis of the axial and radial structure of DC discharges and plasma stratification [16]. It has also appeared in the study of fully-ionized plasma [17].

The diagonal elements of the tensor (7) describe electron diffusion in the configuration space and electron diffusion along the kinetic energy axis with the diffusion coefficient, $D_u = D_r(\boldsymbol{E} \cdot \boldsymbol{E})$, which corresponds to Joule heating. The heating does not depend on the direction of the electric field - it is determined by the absolute value of the electric field *at a given point*. The off-diagonal terms in the diffusion tensor describe additional fluxes in phase space associated with gradients in configuration and velocity space. We aim to clarify the nature of these fluxes under conditions of nonlocal electron kinetics.

Specific source terms $S_0$ for quasi-elastic, inelastic, and Coulomb collisions can be found in [11,17,18]. The quasi-elastic and Coulomb collisions contribute a differential (Fokker-Planck term) to the energy flux $\Gamma$ (see below). Inelastic collisions and ionization processes result in significant energy jumps and change in the number of particles.

## 2. Hybrid Model of Positive Column

Using cylindrical coordinates in configuration space is convenient to describe an axisymmetric positive column plasma inside a cylindrical tube. In an axially homogeneous plasma ($\partial/\partial z = 0$), the electric field can be separated into axial and radial components. These components evolve at different time scales. The axial component, $E_z$, maintaining the plasma is controlled by the external current source. The radial component, $E_r$, is generated by the plasma to support quasi-neutrality. This component evolves at a time scale controlled by the variation in the electron distribution function's isotropic component $f_0$ and by ion transport.

The kinetic equation (6) for the positive column becomes [10]:

$$\frac{\partial f_0}{\partial t} - \frac{1}{r}\frac{\partial}{\partial r}\left[rD_r\left(\frac{\partial f_0}{\partial r} - E_r\frac{\partial f_0}{\partial u}\right)\right] + \frac{1}{v}\frac{\partial}{\partial u}\left[vE_rD_r\left(\frac{\partial f_0}{\partial r} - E_r\frac{\partial f_0}{\partial u}\right)\right] - \frac{1}{v}\frac{\partial}{\partial u}(v\Gamma_u) = S_0^* \tag{8}$$

where the electron flux along the energy axis $\Gamma_u$ is due to electron heating by the axial electric field $E_z(t)$ and energy loss in quasi-elastic collisions that can be described by the Fokker-Planck form, and the term $S_0^*$ describes inelastic collisions. The electron fluxes in phase space $(r, u)$ are:

$$\Phi_r = -D_r\left(\frac{\partial f_0}{\partial r} - E_r\frac{\partial f_0}{\partial u}\right) \tag{9}$$

$$\Gamma = -E_r\Phi_r - D_u\frac{\partial f_0}{\partial u} - V_u f_0 = -E_r\Phi_r - \Gamma_u \tag{10}$$

where $D_u(t) = D_r E_z^2$, $V_u = \delta v u$, and $\delta$ is a fraction of electron energy lost in quasi-elastic collisions of electrons with neutrals.

It is essential to point out that the radial (ambipolar) and the axial (conductive) electric fields play different roles in Eq. (8). The axial electric field $E_z(t)$ contributes to electron heating via the term $D_u\frac{\partial f_0}{\partial u}$. The radial field, $E_r = -\nabla\varphi$, describes electron diffusion in phase space $(r, u)$ over surfaces of constant total energy $\varepsilon = u - \varphi(r, t)$, where $\varphi(r, t)$ is the electrostatic potential.

In the present work, the electric field $E_z(t)$ was calculated as [4]:

$$E_z(t) = \frac{I(t)}{2\pi \int_0^R \mu_e(r,t) n_e(r,t) r dr} \tag{11}$$

where $\mu_e$ is the electron mobility, and $n_e$ is the electron density. The discharge current was prescribed in the form, $I(t) = I_0 \sin \omega t$. As shown below, at low frequencies, the time variation of the electric field $E_z(t)$ can be un-harmonic even for a mono-harmonic variation of the discharge current.

Ions and metastable atoms are described using a drift-diffusion model:

$$\frac{\partial n_i}{\partial t} + \frac{1}{r}\frac{\partial}{\partial r} r \left(\mu_i n E_r - D_i \frac{\partial n_i}{\partial r}\right) = S_i \tag{12}$$

$$\frac{\partial n_m}{\partial t} + \frac{1}{r}\frac{\partial}{\partial r} r \left(-D_m \frac{\partial n_m}{\partial r}\right) = S_m \tag{13}$$

where $\mu_i$ and $D_i$ are the ion mobility and diffusion coefficients of ions, $D_m = D_i$ is the diffusion coefficient of metastables, $S_i$ is the ion production rate, and $S_m$ is the metastable production rate according to the reactions shown in Table 1. It is assumed that gas temperature and ion temperature are $T_i = T_m = T_g = 300$ K.

The set of equations (8-13) satisfies scaling laws [19]: in a given gas, the positive column plasma in a tube radius $R$ depends on $pR$, $I/R$, and $f/p$, where $f = \omega/2\pi$ is the cyclic frequency. Stepwize ionization and volume recombination via binary collisions do not violate the scaling laws, which break in the presence of three-body collisions and substantial deviations from quasi-neutrality.

The classical Schottky model for a positive column assumes a) $S_i = n_e \nu_i(T_e)$, b) $n_i = n_e = n$, c) $\varphi(r) = T_e \ln(n_e)$, which results in the ambipolar diffusion equation for plasma density:

$$\frac{\partial n}{\partial t} - \frac{1}{r}\frac{\partial}{\partial r} r \left(D_a \frac{\partial n}{\partial r}\right) = n\nu_i(T_e) \tag{14}$$

where $D_a = \mu_i T_e$ is the ambipolar diffusion coefficient. This equation determines the value of $T_e$, which does not depend on plasma density and depends only on the reduced electric field $E/p$. The radial distribution of plasma density is described by the Bessel function, $n(r) = n_0 J\left(\frac{\kappa r}{R}\right)$. For the simplest boundary condition, $n(R) = 0$, one obtains $\kappa \approx 2.4$. The plasma density on the axis, $n_0$, is proportional to the discharge current.

We will show that the kinetic model of electrons provides a much richer and more exciting picture of the plasma behavior than the ambipolar Schottky model. We use a simplified three-level model of an argon atom with one metastable state (index $m$), taking into account seven main reactions listed in Table 1. The constants of the electron-induced processes are computed using the corresponding collision cross-sections. The ion mobility and diffusion are $\mu_i = 0.115225 \frac{m^2}{V\,s}\frac{1\,\text{Torr}}{p}$ and $D_i = \frac{\mu_i T_i}{e}$.

*Table 1: The set of reactions for argon plasma (adapted from Ref [20])*

| Name | Reaction | Energy (eV) | Value |
|---|---|---|---|
| Elastic Scattering | $e + Ar \rightarrow e + Ar$ | - | Cross section |
| Excitation | $e + Ar \rightarrow e + Ar_m$ | $\varepsilon_1 = 11.5$ | Cross section |
| Super Elastic | $e + Ar_m \rightarrow e + Ar$ | | Cross section |
| Direct Ionization | $e + Ar \rightarrow 2e + Ar^+$ | $\varepsilon_i = 15.8$ | Cross section |
| Recombination | $e + Ar^+ \rightarrow Ar$ | | $\sigma = 5 \times 10^{-20}$ m$^2$ |
| Stepwise Ionization | $e + Ar_m \rightarrow 2e + Ar^+$ | $\varepsilon_s = 4.3$ | Cross section |
| Penning Ionization | $2Ar_m \rightarrow e + Ar^+ + Ar$ | - | $k_P = 1.2 \times 10^{-15}$ m$^3$s$^{-1}$ |

The boundary conditions for the electron kinetic equation are specified as follows. First, the energy range $u_{max}$ is selected to be sufficiently large to assume $f_0(r, u = u_{max}) = 0$. The boundary condition at $u = 0$

$$\frac{\partial f_0}{\partial r} - E \frac{\partial f_0}{\partial u} \rightarrow 0 \tag{15}$$

corresponds to the absence of electron flux at zero kinetic energy. The boundary condition at $r = R$ is:

$$-\frac{\partial f_0}{\partial r} = \frac{3}{2\lambda} d\Omega f_0 \tag{16}$$

where $d\Omega$ is an effective loss cone [21,22]. In our simulations, we used $d\Omega = 1$.

The boundary conditions for the ion drift-diffusion equation and metastable continuity equation at $r = R$ are:

$$D_i \frac{\partial n_i}{\partial r} + \mu_i E_r n_i = \frac{n_i}{4} \sqrt{\frac{8T_i}{\pi M}} \tag{17}$$

$$D_m \frac{\partial n_m}{\partial r} = \frac{n_m}{4} \sqrt{\frac{8T_m}{\pi M}} \tag{18}$$

The coupled set of the FP kinetic equation for electrons, the drift-diffusion for ions, continuity for metastables, and the Poisson equation for the electric field was solved using COMSOL [23] with an implicit (BDF) time-stepping method. A direct (MUMPS) solver was employed at each time step for all quantities. Logarithmic formulations were used for the Fokker-Planck equation for electrons, the drift-diffusion equation for ions, and the diffusion equation for metastables.

## III.    Results of Simulations

The spatially nonlocal electron kinetics ($f_e(u,r) \neq f_0(u)n_e(r)$) under conditions of temporal nonlocality ($f_e(t,u) \neq f_0(u)n_e(t)$) are analyzed. A range of gas pressures and AC frequencies are chosen to achieve these conditions. The corresponding range of $pR$ for the positive column corresponds to $\lambda < R < \lambda_T$, where $\lambda_T$ is the electron energy relaxation length, a function of electron kinetic energy: $\lambda_T(u) = \lambda(u)/\sqrt{\delta}$ at $u < \varepsilon_1$ and $\lambda_T(u) = \sqrt{\lambda(u)\lambda^*(u)}$ at $u > \varepsilon_1$. Here $\lambda(u)$ is the electron mean free path at kinetic energy $u$, $\delta = \sqrt{2m/M}$ is the fraction of energy transferred in quasi-elastic collisions, and $\lambda^*(u)$ is the electron mean free path for inelastic collisions at the kinetic energy $u$. The AC angular frequency range is $\tau_a^{-1} < \omega < \nu_u$, where $\tau_a = \frac{R^2}{D_a}$ is the ambipolar diffusion time, $D_a = \frac{\mu_i T_e}{e}$, and $\nu_u(u) = \nu(u)\delta + \nu^*(u)$ is the electron energy relaxation frequency.

Figure 1 shows the characteristic frequencies (a) and lengths (b) for Argon pressure of 1 Torr and tube radius of $R = 1$ cm. The bounds on the driving frequency for the dynamic regime are determined by the inverse of the ambipolar diffusion time, $\tau_a^{-1}$ (red), as well as the energy relaxation frequency $\nu_u$. The lower bound on the product $pR$ is determined by the condition that the electron mean free path $\lambda$ (red) is smaller than the tube radius $R$. For the typical electron temperature, $u = \tilde{T}_e = 3$ eV, the cyclic frequency range $f = \omega/2\pi$ corresponding to the dynamic regime is approximately 1 kHz to 10 kHz. The conditions of spatial nonlocality are in the $pR$ range from 0.1 cm Torr to 10 cm Torr.

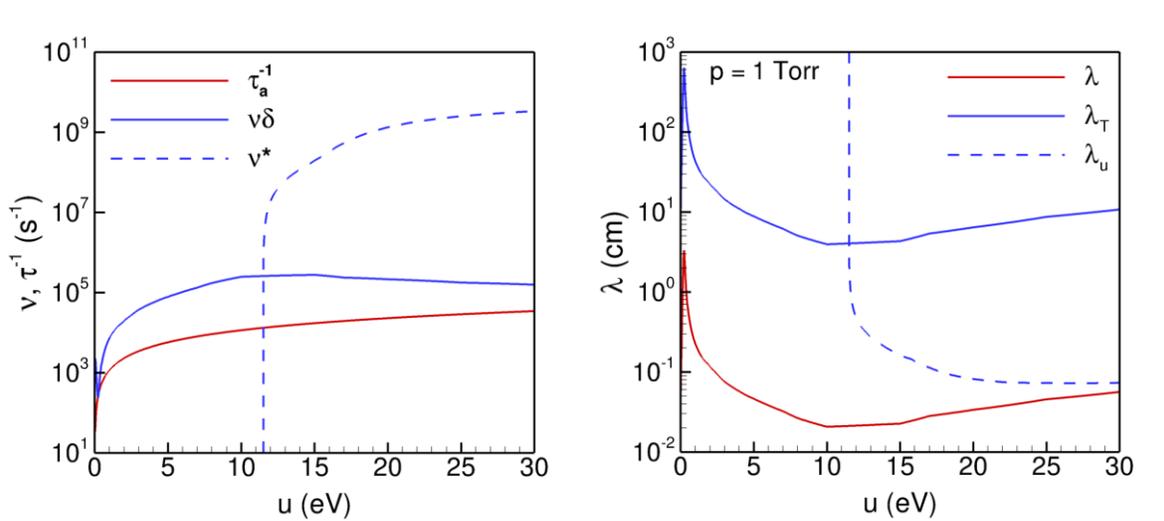

*Figure 1: Characteristic frequencies (a) and lengths (b) as functions of electron kinetic energy in Argon at pressure $p$ of 1 Torr and tube radius $R = 1$ cm.*

1. The base case

All simulations are performed for a radius of $R = 1$ cm and AC with an amplitude $I_0 = 10$ mA. The base case corresponds to the gas pressure of $p = 1$ Torr, and the driving frequency, $f$, in the range from $10^2$ to $10^5$ Hz. Simulations started with a seed electron density of $10^9$ cm$^{-3}$ and allowed to run until reaching a periodic steady state, i.e. $(f_0(t) = f_0(t + 2\pi/\omega))$. After reaching the periodic steady state, the parameters of the positive column are recorded and analyzed.

Figure 2 shows temporal variations of the axial electric field, electron density, and electron temperature on the axis ($r = 0$) for one period of different frequencies. The discharge is in a quasi-static regime at the lowest frequency of 0.1 kHz. In this regime, the electric field and electron temperature remain nearly constant during most of the period, and the plasma density follows in phase with the modulation of the current. Drastic changes of all plasma parameters occur only during a short period when the current passes zero. The electron temperature drops below 0.5 eV during this short time, and the electric field and electron temperature exhibit an overshot. Substantial asymmetry of the temperature variation with time occurs when the electric field crosses zero. The temperature drops smoothly when the field amplitude decreases and increases sharply when the amplitude increases. With increasing frequency, the region of drastic changes near zero current grows relative to the AC period.

At 3 kHz, the discharge operates in the dynamic regime. The amplitude of all oscillations is lower compared to the quasi-static case. There is an extended region where the axial electric field, $E_z$, is close to zero. The asymmetry of the temperature profile remains substantial. Electron heating during the field rise is noticeably faster than the temperature drop during the field decrease. Penning ionization helps maintain plasma when ionization by electron impact becomes small.

At 30 kHz, the driving frequency is substantially above the inverse of the ambipolar diffusion time, so the density oscillations become small. As a result, the electric field oscillations follow the current variations. The changes in electron temperature, ionization, and excitation rates (not shown) remain substantial. As the driving frequency becomes comparable to the collisional relaxation time, the variations of electron temperature decrease, and the average electron temperature is near the value in the quasi-static case ($f = 0.1$ kHz). The observed behavior resembles the previously obtained for ICP using a fluid plasma model [8].

The contribution to the total ionization source rate for all three driving frequencies primarily comprises of stepwise and Penning ionization. Figure *3* shows the time evolution of different ionization and volume loss channels. Stepwise ionization is the main ionization channel over the period. However, Penning ionization becomes essential near zero current. The volumetric recombination is negligible at 1 Torr, and surface recombination is the main channel of particle loss.

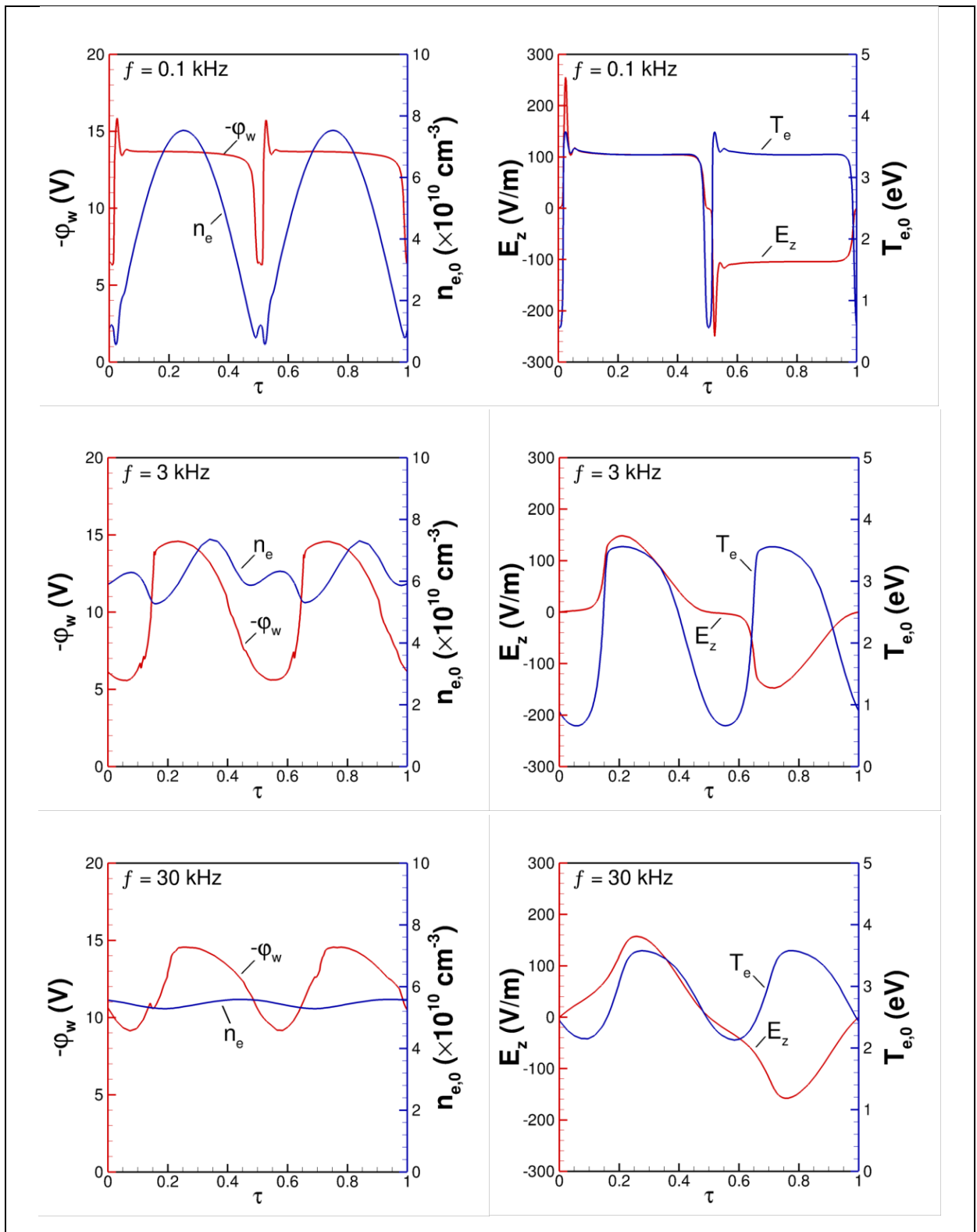

*Figure 2: Temporal variations of the wall potential and electron density (left) and electric field and electron temperature (right) at the tube center for frequencies f = 0.1, 3, and 30 kHz.*

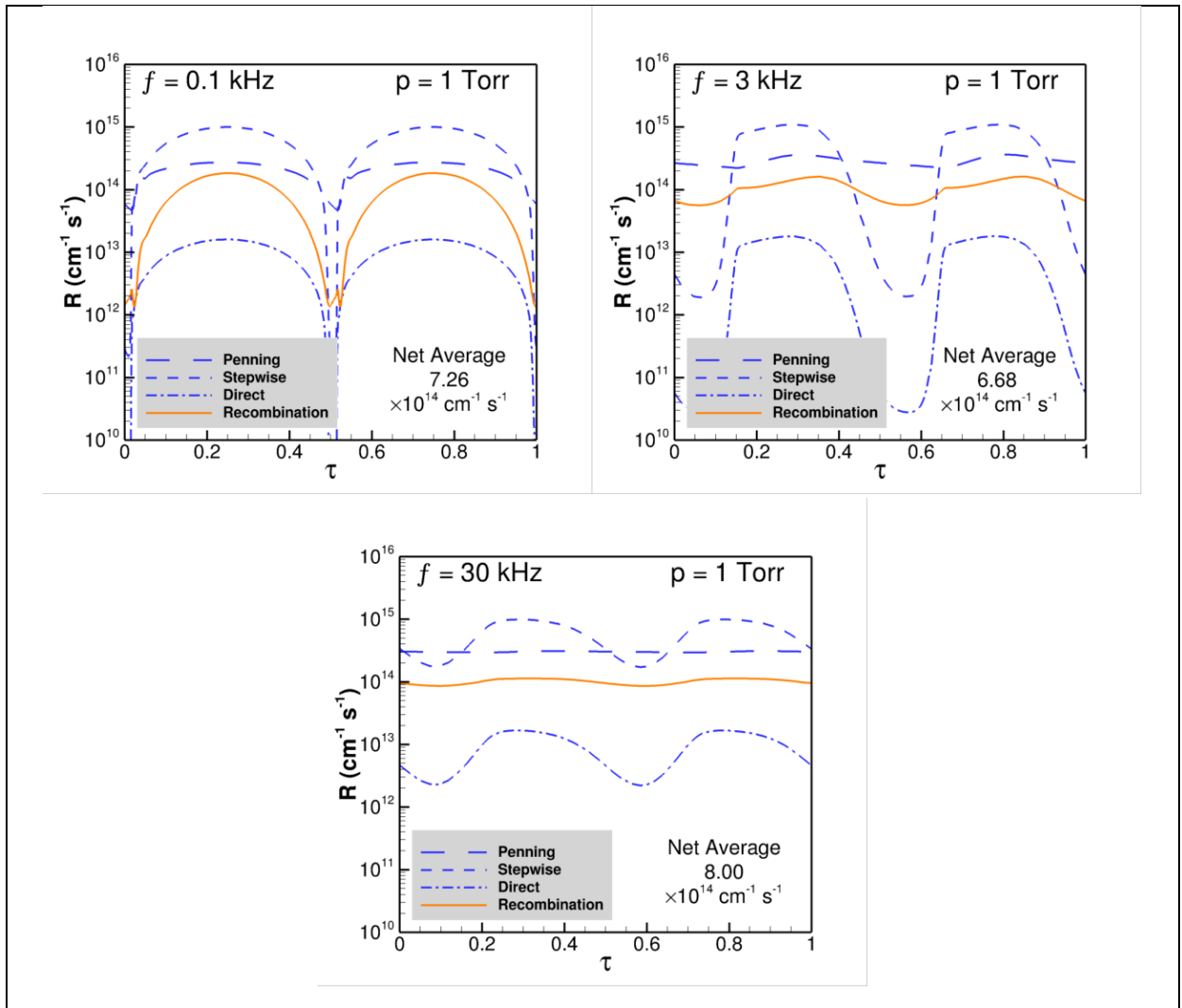

*Figure 3: The rates of Penning, stepwise, and direct ionization as functions of time during the AC period. The loss rate is due to recombination over the entire cross-section.*

The root mean square (rms) electric field, $\bar{E}_z = \sqrt{\langle E_z^2 \rangle}$, electron temperature, $\langle T_e \rangle$, and electron density, $\langle n_e \rangle$, are shown in Figure 4. At driving frequencies below 1 kHz, all three vary relatively weekly with driving frequency. For higher frequencies, a dip in $\langle T_e \rangle$ and a maximum in $\langle n_e \rangle$ occur near 3 kHz. The obtained minimum of $\langle T_e \rangle$ in the dynamic regime agrees with previous experimental observations in discharges driven by modulating currents [2]. The observed minimum of $\langle T_e \rangle$ in [2] occurs at a frequency a factor of 3 to 5 larger than that observed in the present work. This difference can be due to the different masses of Argon and Helium studied in [2].

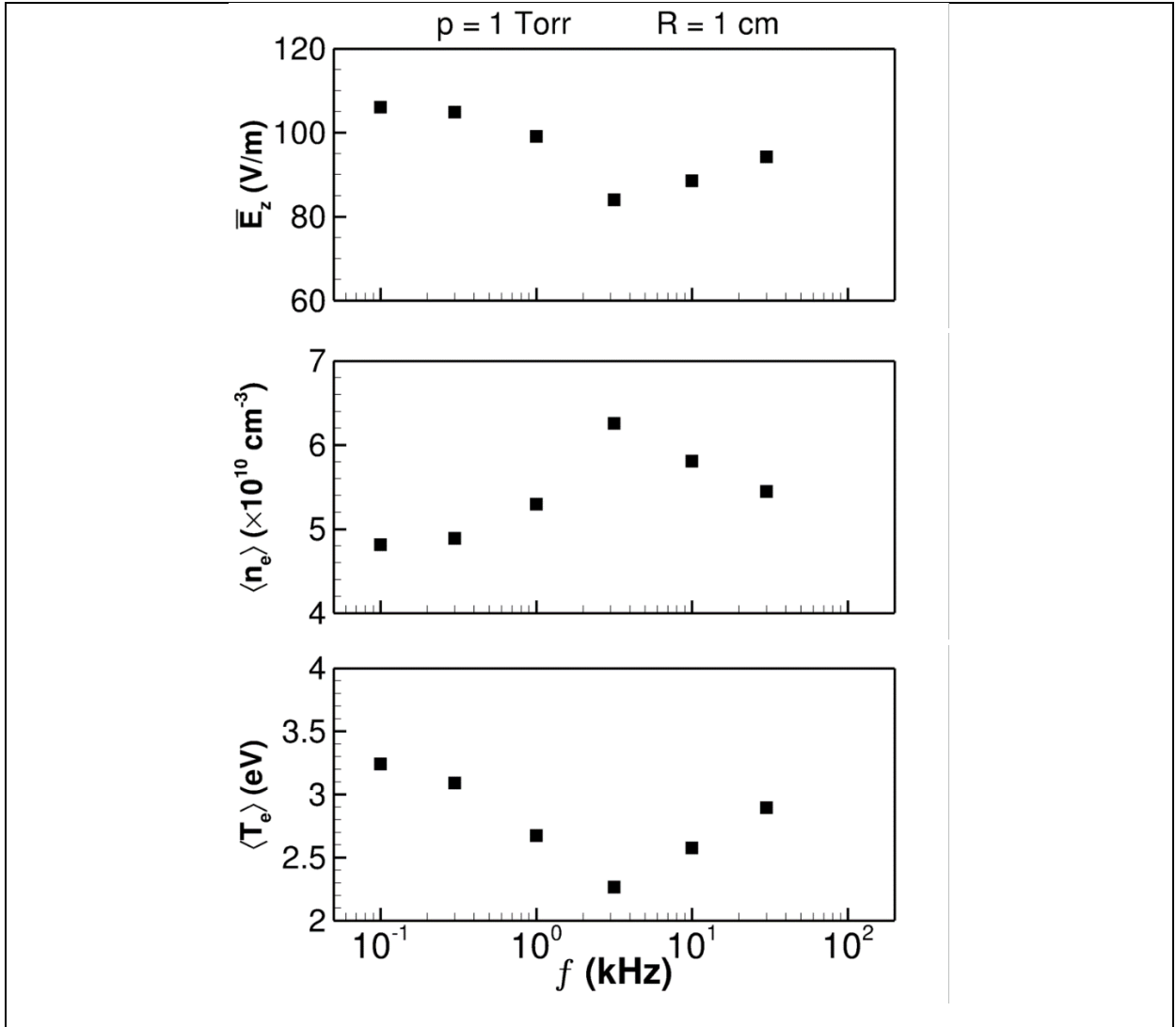

*Figure 4: The dependencies of the rms electric field $\bar{\bar{E}}_z$, electron density $\langle n_e \rangle$, and electron temperature $\langle T_e \rangle$ at the tube axis on driving frequency $f$, for $p = 1$ Torr.*

Figure 5, Figure 6, and Figure 7 show the variations of the EEPF, $f_0(u,t)$, on the tube axis at frequencies 0.1, 3, and 30 kHz. At low frequency, 0.1 kHz, the shape of the EEPF remains approximately the same for most of the period, except when the current crosses zero. When the current crosses zero, the EEPF shape changes sharply (Figure 5).

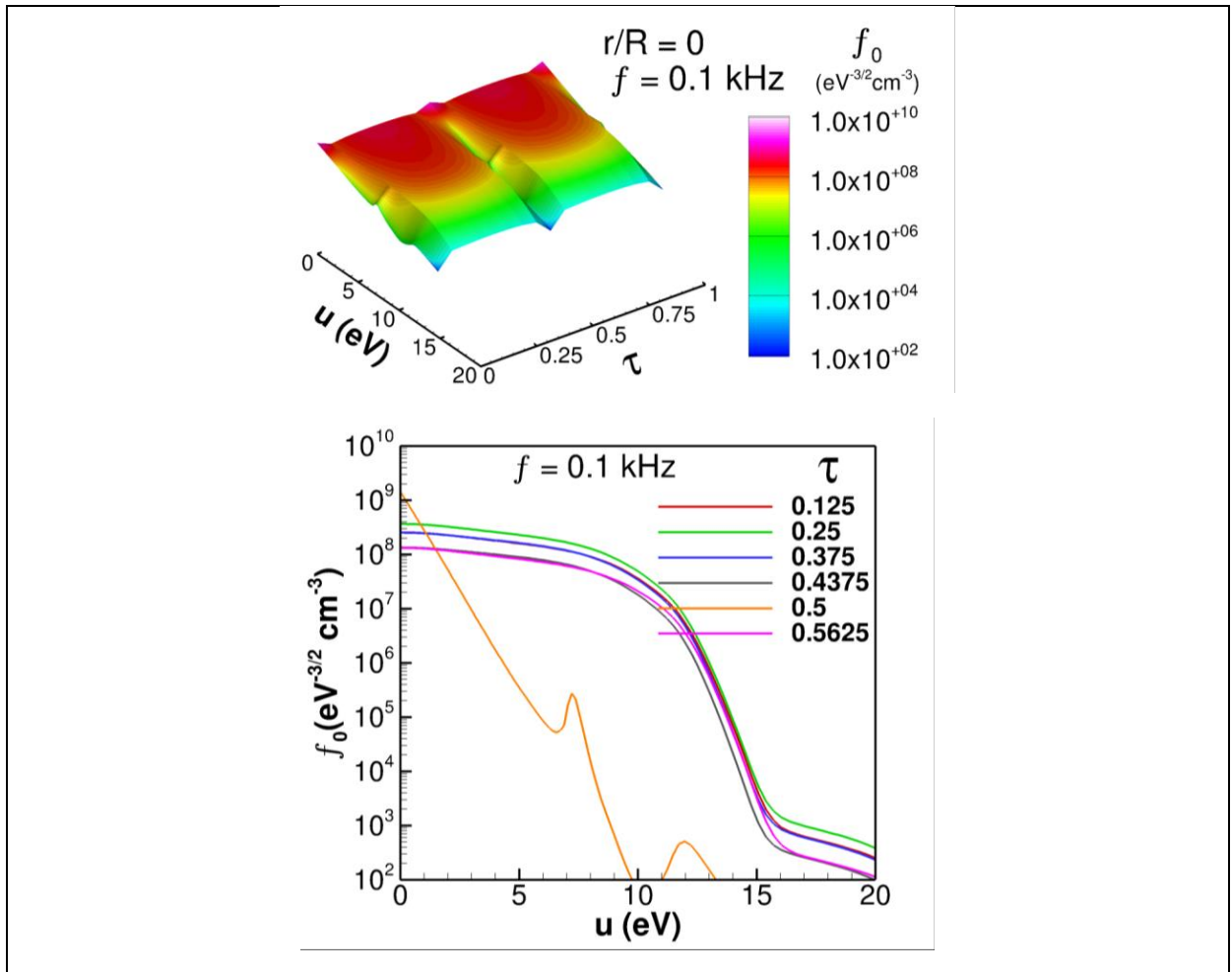

*Figure 5: The EEPF, $f_0(u)$, as function of kinetic energy six times during an AC period at $f = 0.1$ kHz.*

At 3 kHz, the shape of the EEPF varies substantially over the period (Figure 6). At low currents, the EEPF becomes non-monotonic. Unlike 0.1 kHz, this non-monotonic shape remains for most of the period. The first of the two local maximums is a result of Penning ionization. The second is a result of super-elastic collisions. Both of these processes occur as the number density of metastables remains substantial during the portion of the period where electron temperature becomes low.

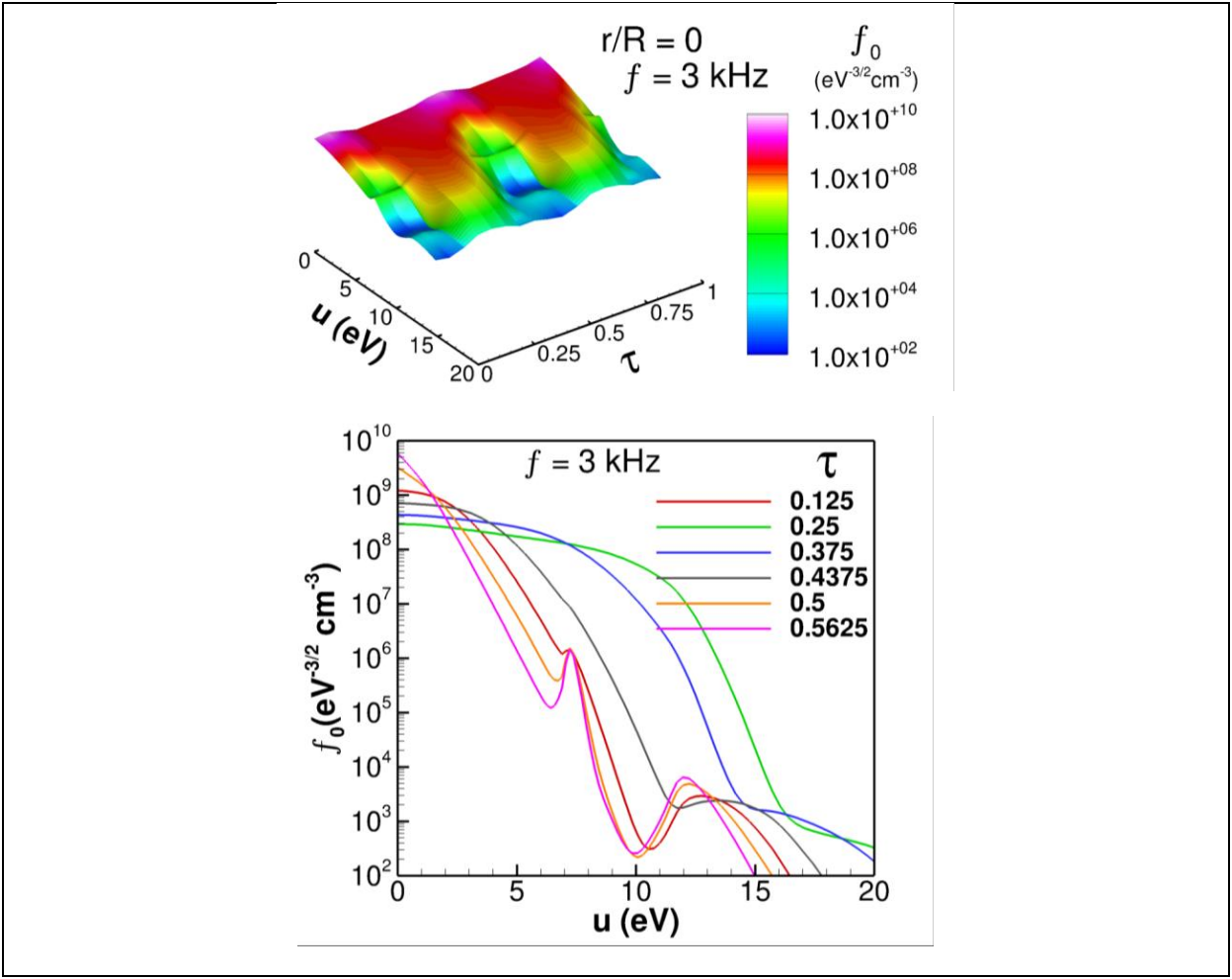

*Figure 6: The EEPF, $f_0(u)$, vs kinetic energy at six times during one AC period at $f = 3$ kHz.*

Finally, at 30 kHz, the variations of the EEPF shape are significantly reduced (Figure 7). At this frequency, the electron number density varies slightly, but the variations in electron temperature remain substantial. The EEPF is a monotonic function of the kinetic energy at 30 kHz.

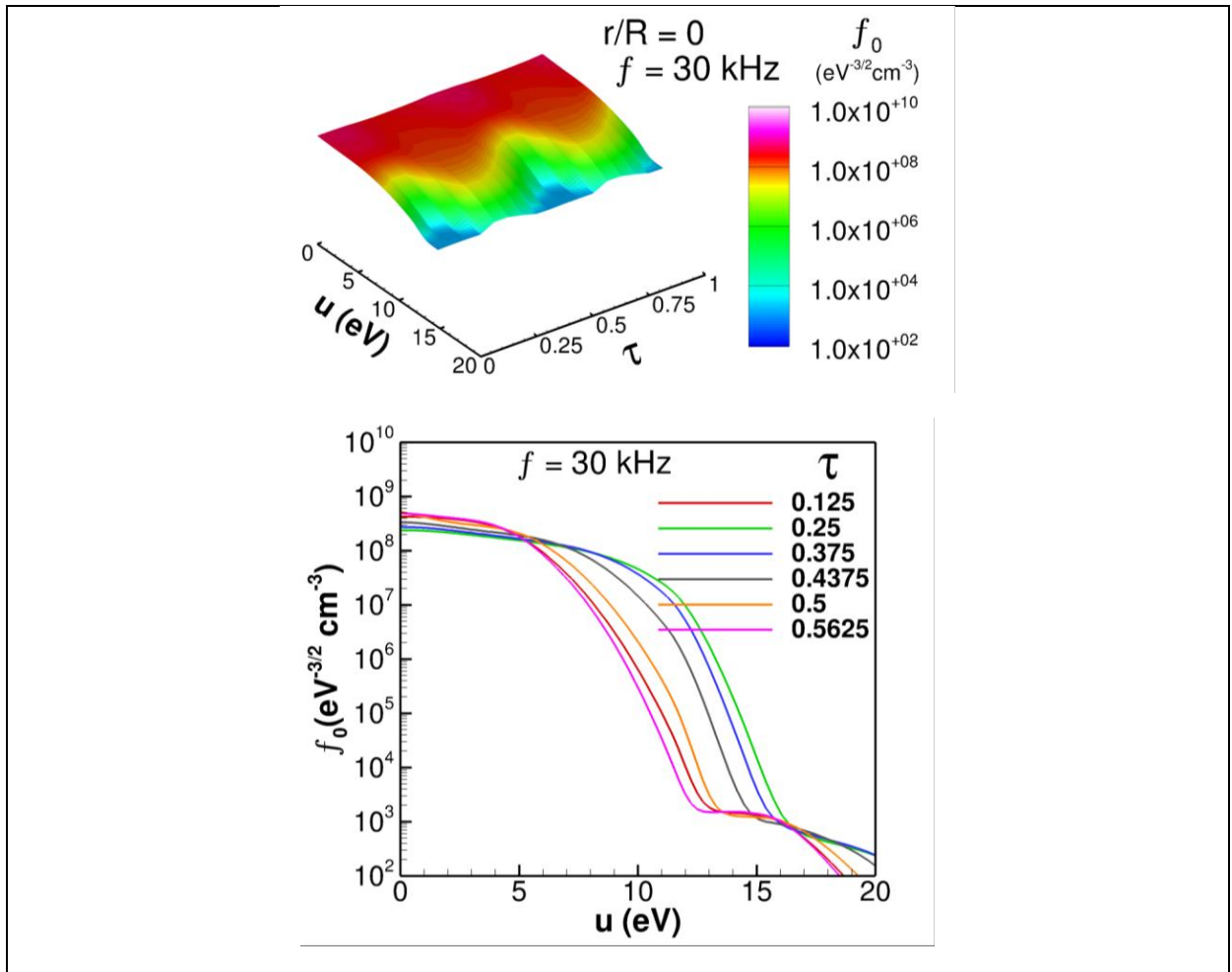

*Figure 7: The EEPF, $f_0(u)$, vs kinetic energy at four times during a half AC period at 30 kHz.*

## 2. The pressure dependence

We have conducted simulations at $f = 3$ kHz for different $p$ in the range $0.1 – 10$ Torr. Figure 8-10 show the radial distributions of electron density (a), metastable density (b), electron temperature (c), and ionization rate (d) over one period for pressures of 3, 6, and 10 Torr. At $p = 3$ Torr, the radial distributions of electron and metastable densities, electron temperature, and ionization rate are monotonic. Moderate radial constriction occurs for the metastable density due to stepwise ionization (dotted line in Figure 8d).

At $p = 6$ Torr, an off-axis peak of plasma density forms (Figure 9). Steep off-axis maxima of metastable density and stepwise ionization rate are also present. The maximum plasma density peak relative to the tube center occurs near the maximum current ($\tau = 0.25$). At zero current ($\tau = 0.5$), the plasma density nears, but does not become, monotonic.

At $p = 10$ Torr, the plasma density peak is located closer to the wall and is more pronounced (Figure 10). A pronounced off-axis maximum plasma density (~ 20% greater than the axial value) remains throughout the period. Additionally, a "spike" in electron temperature near the wall occurs near the current zero. This spike is most pronounced at the highest pressure ($p = 10$ Torr) studied in the present paper.

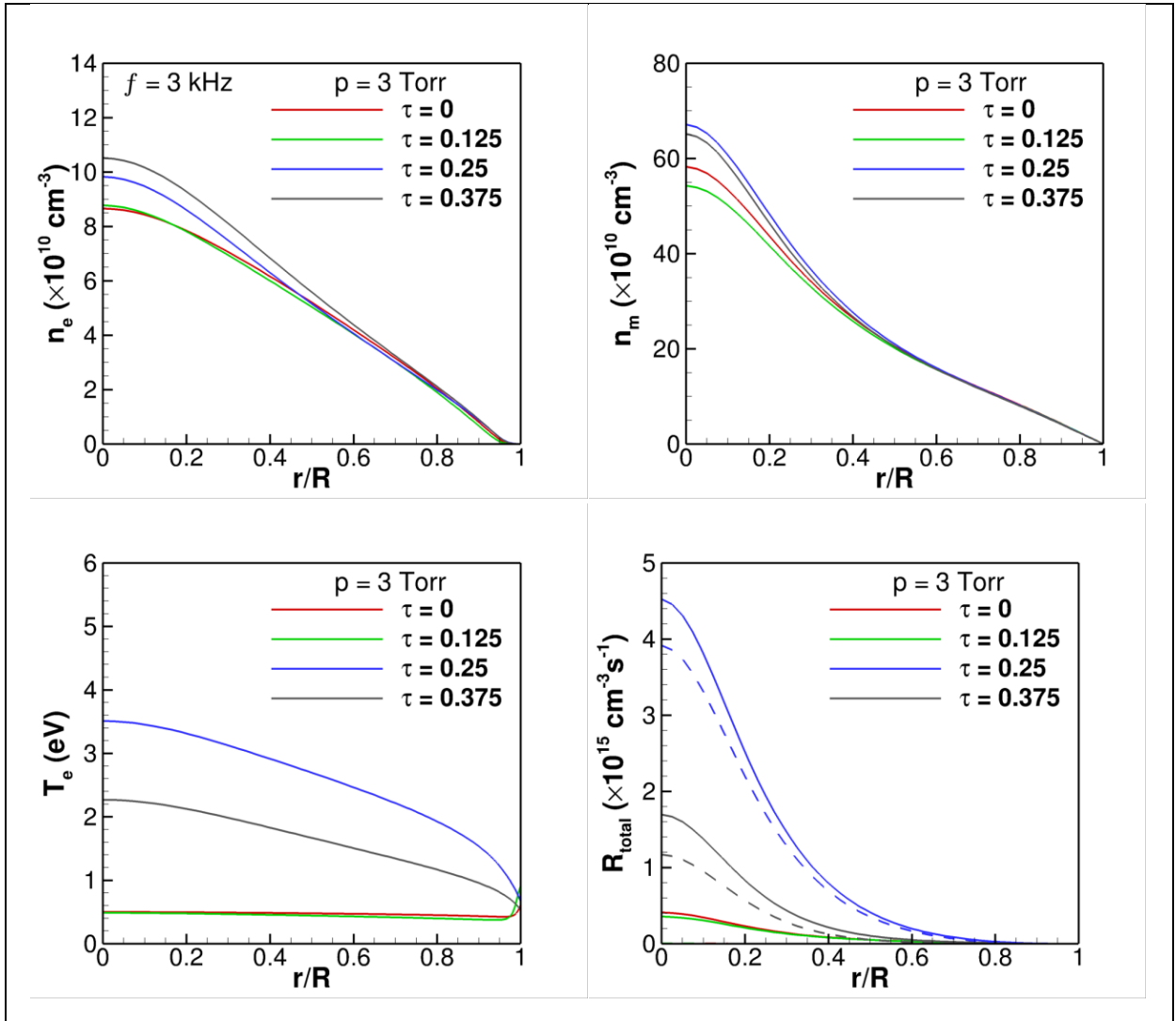

Figure 8: Radial distributions of electron density (a), metastable densities (b), electron temperature (c), and ionization rates (d) four times throughout half a period for pressure $p = 3$ Torr. Dashed lines in the distribution of ionization rates show the contribution from stepwise ionization.

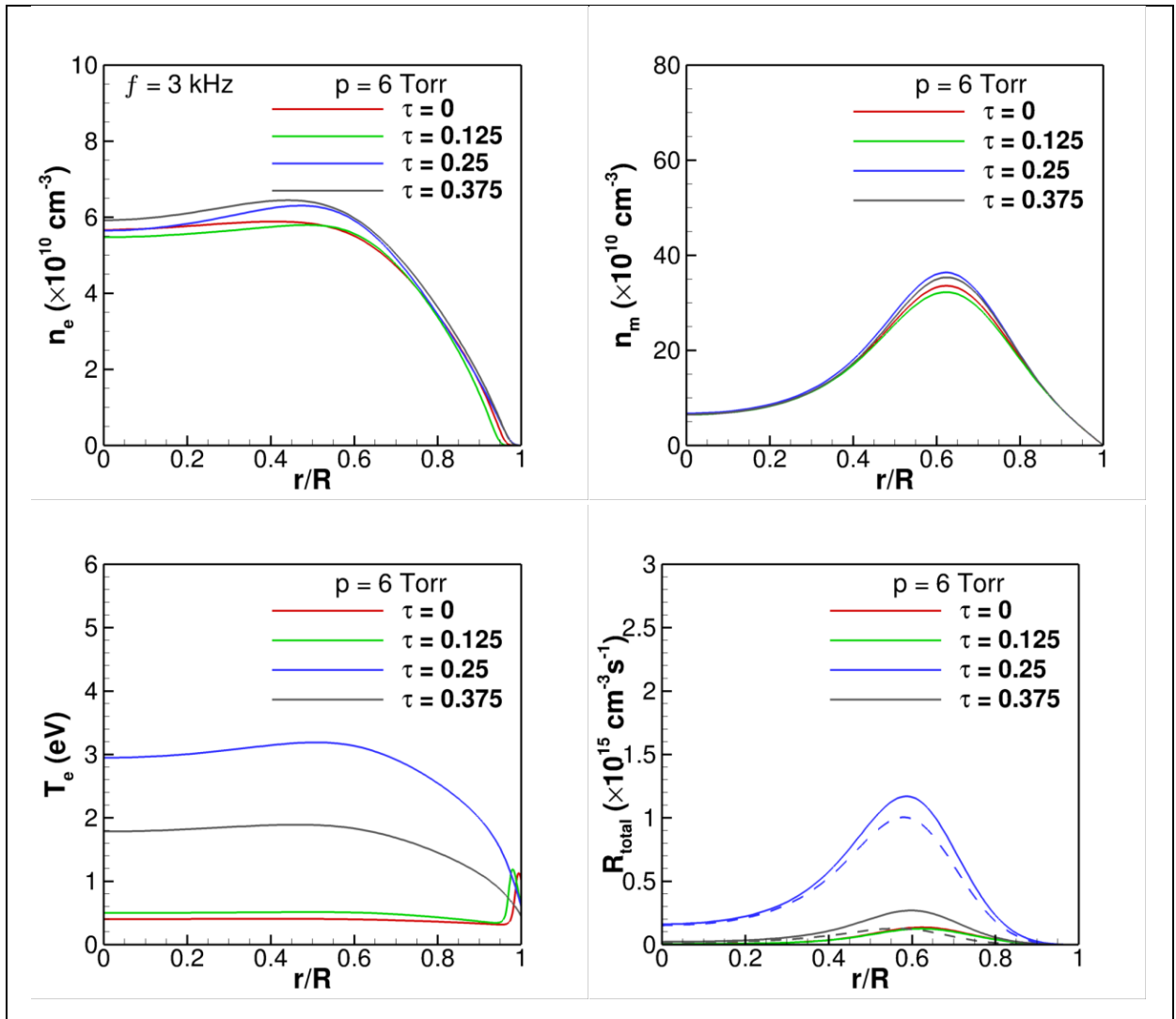

*Figure 9: Radial distributions of electron density (a), metastable densities (b), electron temperature (c), and ionization rates (d) for four times throughout half a period at p = 6 Torr. Dashed lines in the distribution of ionization rates show the contribution from stepwise ionization.*

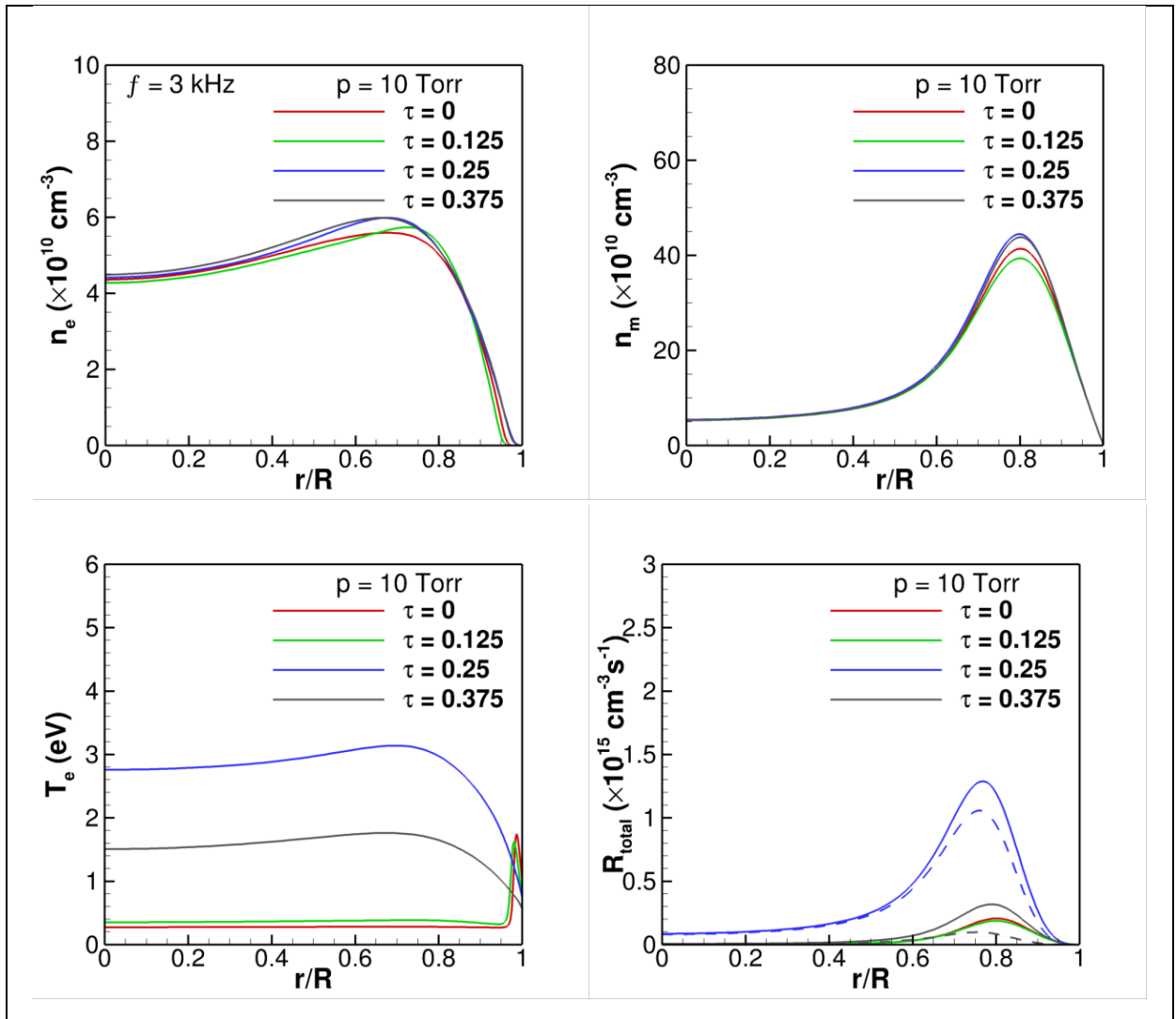

*Figure 10: Radial distributions of electron density (a), metastable densities (b), electron temperature (c), and ionization rates (d) four times throughout half a period at p = 10 Torr. Dashed lines in the distribution of ionization rates show the contribution from stepwise ionization.*

Figure 11 compares the dependence of the reduced electric field $E/p$ on the reduced frequency $f/p$ for varying pressure at $f$ = 3 kHz (in red) and varying frequency at $p$ = 1 Torr (in green). The $f/p$ scaling is not satisfied in our model. A much stronger dependency on pressure observed in our simulations may be due to Penning ionization. Further studies of the $f/p$ scaling could be the subject of additional studies.

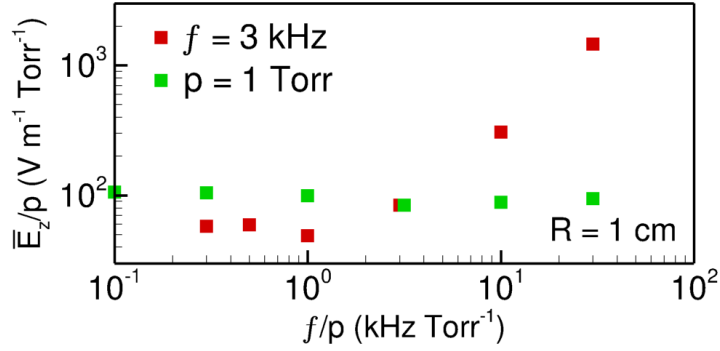

*Figure 11: Reduced electric field E/p vs. f/p by varying pressure at 3 kHz (in red) and varying frequency at 1 Torr (in green).*

## IV. Peculiarities of the Dynamic Regime of Discharge Operation

The dynamic regime of plasma operation becomes possible due to the significant difference in the ambipolar diffusion time, $\tau_a$, and the electron energy relaxation time, $\tau_u$. This disparity of time scales is a consequence of the small electron mass compared to the ion mass. In our studies, $\tau_u$ is shorter than $\tau_a$ by two orders of magnitude. The peculiarities of plasma dynamics in the frequency range $\tau_u \ll \omega^{-1} \ll \tau_a$ have been previously studied in Ref. [8] using a fluid model for electrons. In the dynamic regime, the plasma density varies slightly over the AC period, but the electron temperature, ionization rate, and the EEDF shape change significantly. This operating regime has found application in fluorescent lamps driven at 20 –100 kHz [24].

### 1. Increased efficiency of plasma maintenance in the dynamic regime

Figure 12 illustrates the origin of the increased efficiency of plasma maintenance in the dynamic regime. In a positive column, the number of ionizations must be equal to the number of electron losses over the AC period:

$$\int_0^T \nu_i(t)dt = \int_0^T \tau_a^{-1}(t)dt \qquad (19)$$

This condition defines the electric field and the mean electron energy $u_0$ required to maintain the plasma. In particular, Eq. (19) defines a specific value of the electric field $E_0$ in DC discharges.

In AC discharges, the axial electric field changes sign oscillating during the current period. However, substantial modulations of the mean electron energy (temperature) occur only in a limited frequency range corresponding to the dynamic regime. At high frequencies, $f \gg \tau_u^{-1}$, modulations of electron temperature vanish, and the mean value of the temperature depends on the effective value of the electric field, $E_0/\sqrt{2}$, defined by Eq. (19).

Due to the nonlinear dependence of the ionization rate, $\nu_i(u)$, on the electron energy (see Figure 12), time modulations of the electron kinetic energy produce more ionizations during the period than required to balance losses that scale linearly with electron kinetic energy. Thus, to maintain the balance, the rms electric field, and the mean electron temperature must be smaller, and the efficiency of plasma maintenance must be higher in the dynamic regime.

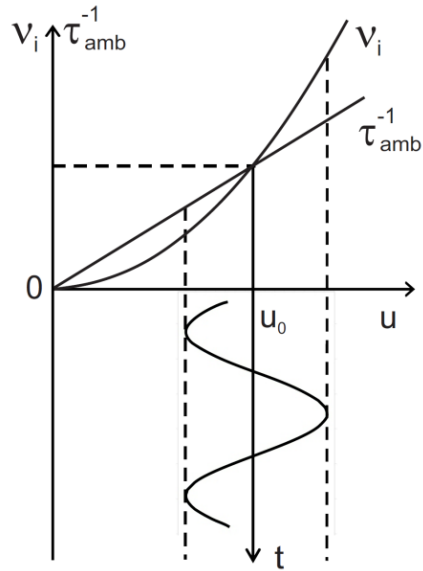

*Figure 12: Illustration of enhanced ionization by the time modulation of the electric field due to nonlinear dependence of the ionization frequency on the field strength (adapted from Ref. [3])*

### 2. Nonlocal kinetic effects in the dynamic regime

Several assumptions of the fluid model discussed above become invalid when nonlocal effects are essential. First, the EEDF cannot be factorized as $f_e(u,r) \neq f_0(u)n_e(r)$, which can be termed *spatial* nonlocality. Second, the EEDF cannot be factorized as $f_e(u,t) \neq f_0(u)n_e(t)$, which can be termed *temporal* nonlocality. In low-pressure AC discharges operating in dynamic regimes, both spatial and temporal nonlocality play a role $f_e(u,r,t) \neq f_0(u)n_e(r,t)$. As illustrated in the present paper, nonlocal electron kinetics produces exotic effects in the dynamic regime.

As discussed in [1], the energy relaxation times and lengths are strikingly different for the EDF body and tail. For the EDF tail ($u > \varepsilon_1$) the characteristic time is $1/\nu^*$ and the energy relaxation length can be estimated as

$$\lambda_u^* = \sqrt{D_r/\nu^*} = \sqrt{\lambda \lambda^*} \sim (3-10)\,\lambda \qquad (20)$$

The characteristic time $\tau_f$ of the EEDF body relaxation at $u < \varepsilon_1$, depends on the overall energy balance of electrons. At elevated gas pressures ($pR > 1$ cm Torr), energy loss in quasi-elastic collisions controls the electron energy balance, $\tau_f \sim \tau_u = \frac{1}{\delta \nu}$. This time of the EEDF body relaxation is much longer than the corresponding time for the tail relaxation.

At low gas pressures ($pR < 1$ cm Torr), when inelastic collisions control the energy balance of electrons, the characteristic time $\tau_f$ of the EEDF's body relaxation is much longer than the time

$\tau_E = \frac{1}{\nu_E} = \varepsilon_1^2/D_E$, that is required for electrons to diffuse from zero energy to the excitation energy $\varepsilon_1$. The corresponding length is

$$\lambda_E = \sqrt{D_r \tau_E} = \varepsilon_1/(eE) \tag{21}$$

During the time $\tau_E$, quasi-elastic energy losses are small because $\tau_E < \tau_u$. The asymmetry of the electron temperature in AC discharges during the transit over zero current is a consequence of different cooling and heating rates. When the amplitude of the electric field decreases, the EEDF relaxation occurs slowly over the time $\tau_f \sim \frac{1}{\delta \nu}$ (this prevents the electron temperature from dropping to the gas temperature). When the amplitude of the electric field decreases, the EEDF formation occurs much faster with the characteristic time $\tau_E < \frac{1}{\delta \nu} = \tau_u$. This asymmetry was previously observed in simulations in Neon [25] and was discussed in the book [1] for spatially uniform plasma.

### 3. The total energy viewpoint

Peculiarities of nonlocal electron kinetics in the dynamic regime can be best illustrated using the *total* energy approach. Using total energy $\varepsilon = u - \varphi(r,t)$ as an independent variable, we can rewrite Eq. (8) in the form:

$$\frac{\partial f_0}{\partial t} - \frac{\partial \varphi}{\partial t}\frac{\partial f_0}{\partial \varepsilon} - \frac{1}{vr}\frac{\partial}{\partial r}\left(rvD_r \frac{\partial f_0}{\partial r}\right) - \frac{1}{v}\frac{\partial}{\partial \varepsilon}(v\Gamma_u) = S_0^* \tag{22}$$

The speed $v(r, \varepsilon(t))$ becomes the Lame coefficient in the $(r, \varepsilon)$ space.

The transition to the total energy corresponds to the diagonalization of the diffusion tensor (7). Although the transition to total energy simplifies the kinetic equation (8) by removing the cross-derivative terms, it makes the integration domain $v(r, \varepsilon, t) > 0$ for Eq. (19) more complicated and time-dependent [26]. Eq. (8) is preferable for the numerical solution, as discussed in [16]. However, the total energy viewpoint is more transparent in understanding the simulation results.

Many publications have used the total energy approach to analyze electron kinetics in gas discharges (see [27,28] and references therein). Eq. (22) coincides with the heat conduction equation in a thermo-isolated flexible bag bounded by the curve $v(r, \varepsilon, t) = 0$ and $\varepsilon = \varphi_w$ (the top of the bag). The heat can escape to the wall ($r=R$) only above the bag top, at $\varepsilon > \varphi_w$. The second term in Eq. (22) describes the effect of the bag shape changes. Electron production in the ionization events corresponds to the heat release at the bottom of the bag. Excitations of the atomic levels by electron impact produce a heat sink at the top of the bag, and the corresponding heat release at the bottom, as described by the right-hand side of (22). The excitations do not change the total heat amount in the bag as the hot electrons disappear, and the same number of cold electrons reappear at the bottom.

The heat conduction analogy helps illustrate different scenarios of the EEDF formation depending on $pR$ values. At low values of $pR$, when the electron energy loss in elastic collisions is negligible, the EEDF of trapped electrons (with $\varepsilon < \varphi_w$) depends only on the total electron energy and does not depend explicitly on the spatial coordinate $r$ [29]. Figure 13 demonstrates that the total energy scaling is perfectly satisfied in our simulations at low pressure, whereas the EEPF depends explicitly on the radial position at high pressure. Indeed, at $p = 0.1$ Torr (Figure 13a), the relaxation length is sufficiently large compared to the tube radius for both low and high-energy electrons. The EEPF body *and* tail are functions of total electron energy. With increasing $pR$, the explicit dependence of the radial position first appears in the tail, as was observed in our simulations at 1 Torr (see the base case above). With a further increase of $pR$, the entire body starts depending on the kinetic rather than total energy. Figure 13b illustrates this effect for $p = 10$ Torr.

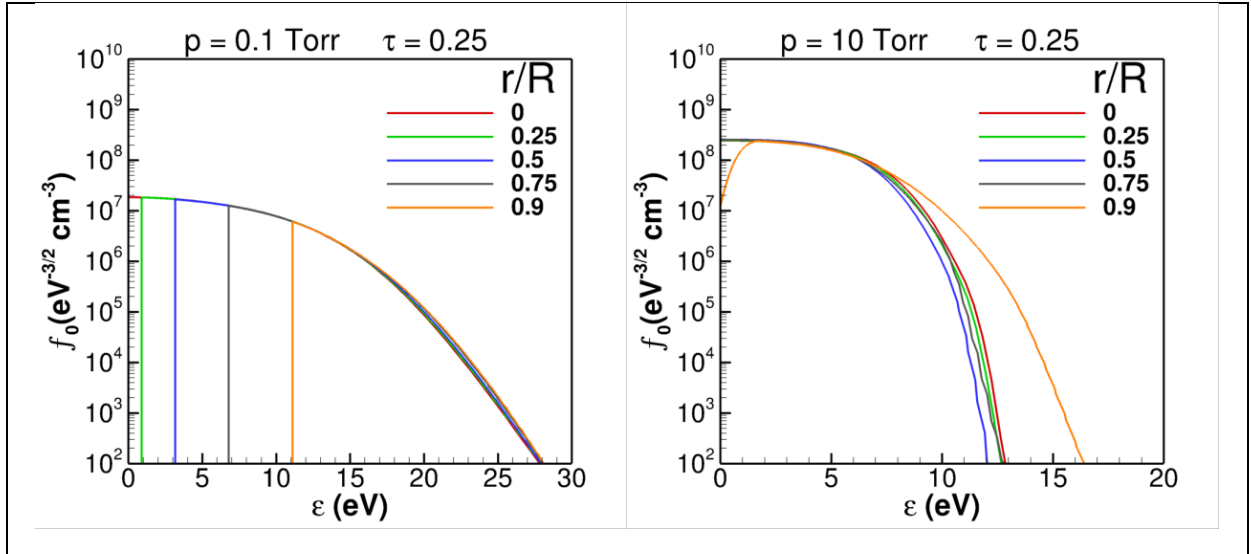

Figure 13: The EEPF at different radial positions as a function of total energy for 0.1 Torr (a) and 10 Torr (b) at a peak current ($\tau = 0.25$) for $f = 3$ kHz.

Additional deviations from the total energy behavior can occur in AC discharges when the first two terms in Eq. (22) become essential. Figure 14 shows the EEPF plotted as a function of the total energy at the times of maximum and zero currents. For 0.1 kHz (top part of Figure 14), the body of EEPF is a function of the total energy, but its tail depends explicitly on the radial coordinate (because $\lambda_u^* < R$ for $pR \sim 1$). The tail is most depleted by the fast electrons near the axis due to the loss of these electrons in inelastic collisions. Therefore, electron flux in phase space (see the section below) must be directed toward the axis at these energies. Near zero current ($\tau = 0.5$), the body is a function of the total energy except for the two peaks, which depend on the radial position. The first peak at the energy of about 7 eV is due to the electrons generated by Penning ionization. The second peak at the energy of about 12 eV is associated with super-elastic collisions and is proportional to the low-energy part (body) of the EEPF [30].

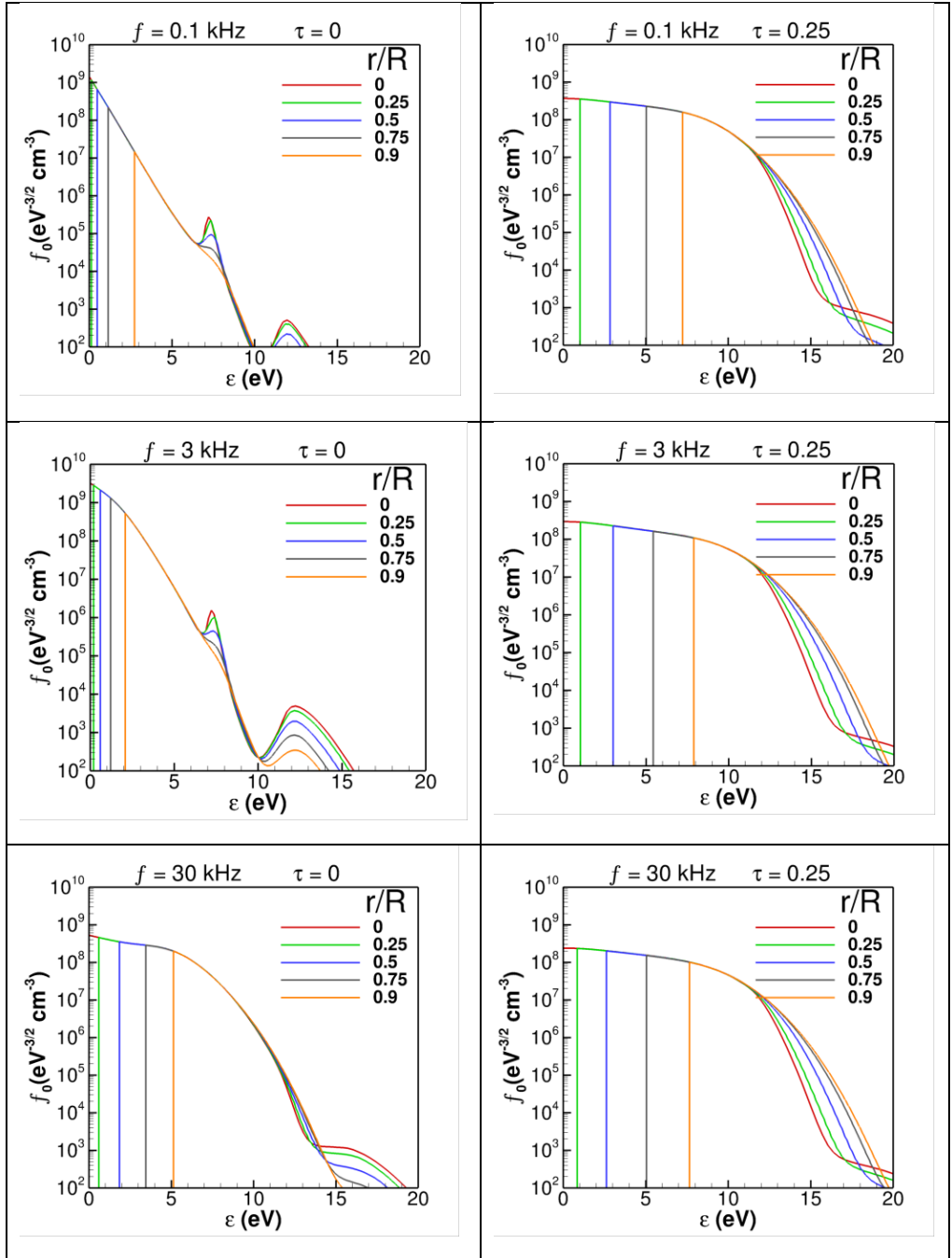

*Figure 14: The EEPF as a function of total energy, $f_0(\varepsilon)$, at maximum ($\tau = 0.25$) and zero ($\tau = 0.5$) currents for $f = 0.1, 3, 30$ kHz*

At $f = 3$ kHz (middle part of Figure 14), the picture is qualitatively similar. Throughout the period, explicit radial dependence occurs only at high electron energies (above ~10 eV). At $f = 30$ kHz (bottom part of Figure 14), the EEPF at maximum current looks similar to the previous cases, but the EEPF near current zero lacks the peaks present at lower driving frequencies. The electrons have insufficient time to cool down at this frequency when the electric field $E_z(t)$ crosses zero.

## 4. Electron fluxes in phase space

The formation of electron fluxes in phase space for a DC positive column under nonlocal conditions has been discussed in several publications [21,22,29,7]. This section discusses the peculiarities of electron fluxes in phase space in the dynamic regime. The kinetic equation (14) can also be written as:

$$\frac{\partial f_0}{\partial t} - \frac{\partial \varphi}{\partial t}\frac{\partial f_0}{\partial \varepsilon} + \frac{1}{vr}\frac{\partial}{\partial r}(rv\Phi_r) + \frac{1}{v}\frac{\partial}{\partial \varepsilon}(v\Gamma) = S_0 \qquad (23)$$

The electron fluxes in $(r, \varepsilon)$ phase space are simplified compared to fluxes (4,5) in $(r, u)$ phase space:

$$\Phi_r = -D_r \frac{\partial f_0}{\partial r} \qquad (24)$$

$$\Gamma = -\Gamma_u \qquad (25)$$

The flux $\Phi_r$ must be zero on the axis where $\frac{\partial f_0}{\partial r} = 0$ and also at the boundary $v(r, \varepsilon, t) = 0$, which reflects electrons. The flux $\Gamma$ consists of the diffusion (heating) component directed upwards and the convection (cooling) component directed downward.

Figure 15 shows an example of flux distributions in phase space for different driving frequencies at three times during the period for 1 Torr. Under these conditions, the energy loss in the elastic collisions is small, and the flux $\Gamma$ is directed upwards during most of the AC period. At times $\tau = 0.25, 0.375$, the exception to this is near the tube center at energies about $\varepsilon/\varepsilon_1 \approx 0.7$. Here, a net downward flux in energy results in the vortex formation in phase space. The radial flux of fast electrons is directed toward the axis, which is consistent with Fig 12, where the number of fast electrons is depleted near the center due to inelastic collisions because $\lambda_u^* < R$ for $pR \sim 1$.

As seen in Figure 2, the wall potential drops below the excitation energy threshold $\varepsilon_1$ when current is zero ($\tau = 0.5$). The energy fluxes at this time (Figure 15, bottom row) are solely from losses in elastic collisions (via $V_u$ term). At the two smaller frequencies, 0.1 and 3 kHz, the electric potential at the wall drops far below $\varepsilon_1$, and a strong radial flux is formed across the tube due to escape of most energetic electrons to the wall. This flux is due to the diffusive cooling of electrons with total energy $\varepsilon$ above the wall potential.

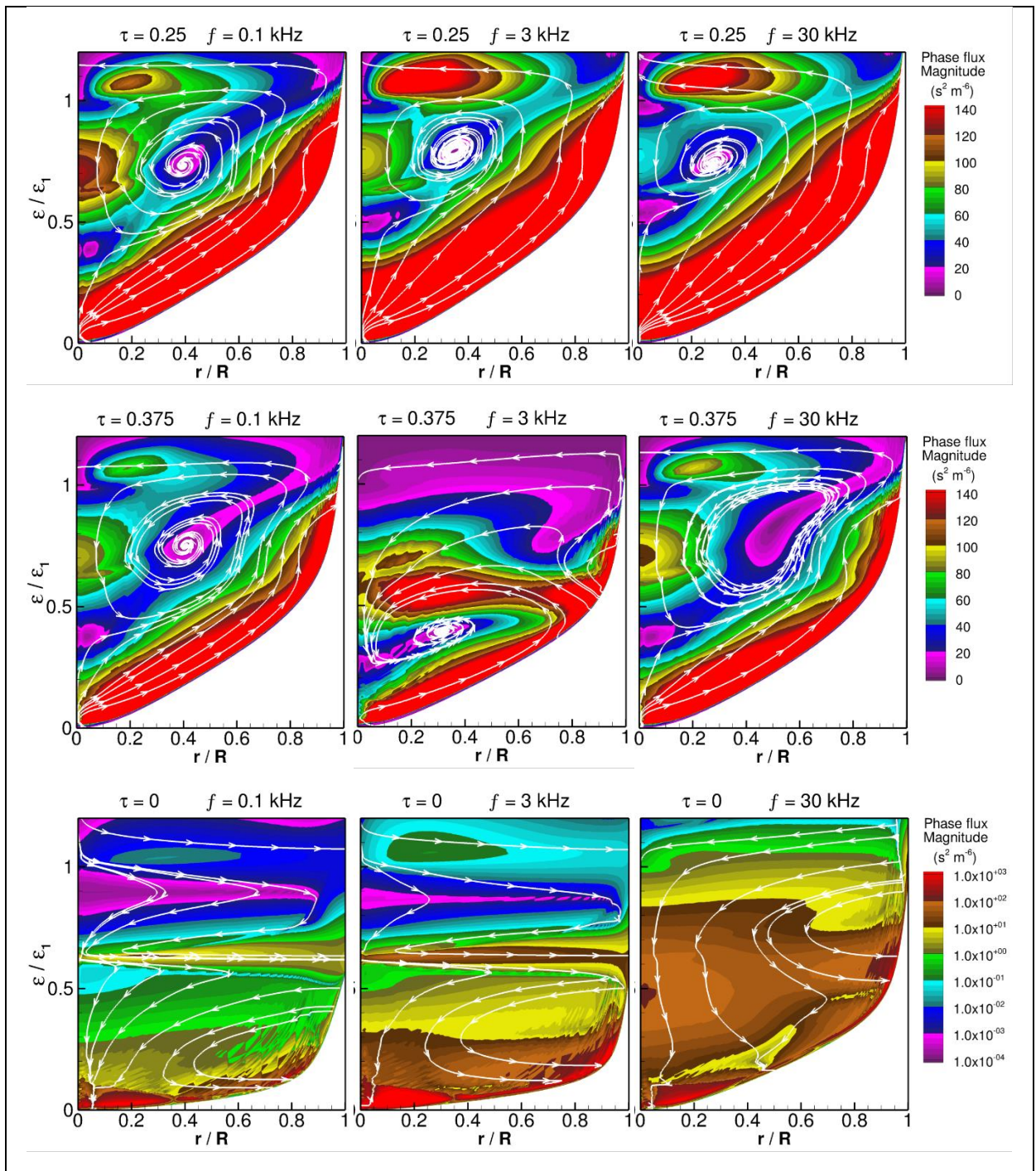

*Figure 15: Electron flux magnitude (color) and streamlines in phase space at times $\tau = 0.25$ (top row), 0.375 (middle row), and 0 (bottom row) for $f = 0.1$ (left), 3 (center), and 30 (right column) kHz.*

## V.     Conclusions

We have performed hybrid kinetic-fluid simulations of a positive column in AC Argon discharges in the collisional regimes ($\lambda \ll R$) over various driving frequencies and gas pressures. We have focused on the low $pR$ conditions ($\lambda_T \gtrsim R$) when the spatial nonlocality of the EEDF is substantial. It was confirmed that the most efficient (minimal power) conditions of plasma maintenance are observed in the dynamic regime. The minimal values of a period-averaged axial electric field and the electron temperature have been observed at $f/p$ values of about 3 kHz/Torr. The ionization rate and plasma density reached maximal values under these conditions.

The Fokker-Planck kinetic solver for EEDF was used to properly account for the nonlocal kinetic effects associated with spatial and temporal nonlocality of the EEDF. We obtained an efficient solution of the anisotropic tensor diffusion equation in phase space using kinetic energy as an independent variable. We clarified the role of the diagonal terms, which describe space diffusion and electron heating in phase space, and off-diagonal terms, which are responsible for the energy conservation during electron diffusion in phase space (over surfaces of constant total energy). An important conclusion worth emphasizing is that electron heating depends on the absolute value of the local electric field. As a result, the ambipolar electric field, which traps electrons, contributes to electron heating on the plasma periphery (at $r \sim R$), where the radial electric field exceeds the axial field.

We have shown that the kinetic view uncovers a more exciting and rich physics than the classical ambipolar diffusion (Schottky) mode. The electron fluxes in phase space illustrate an interesting behavior under conditions of temporal and spatial nonlocality of the collisional electron transport. The electrons at different energies can flow in opposite directions. The energy component of the flux changes its direction during the AC period. The difference between the heating and cooling rates results in the asymmetry of the EEDF behavior and the time variation of electron temperature when the current passes through zero value.

We have confirmed that electron diffusion with conservation of the total electron energy is valid when the temporal variations are slow compared to the EEDF relaxation time. The EEDF under these quasi-static conditions is a function of the total energy only and depends on space and time implicitly. Deviations from this behavior have been detected during rapid variations of the electric fields near current zero when the additional transient terms in the kinetic equation become essential. The radially non-monotonic distributions of excitation rates, metastable densities, and even the plasma density have been observed in our simulations with increasing $pR$. The off-axis peak of the plasma density in the dynamic regime predicted by our model has never been observed in experiments. We hope our results stimulate further experimental studies of the AC positive column. The kinetic analysis could help uncover new physics even for such a well-known plasma object as a positive column in noble gases.

The Knudsen regimes ($\lambda \geq R$) of the weakly collisional positive column will be the subject of future studies.

## Acknowledgments

This work was supported by the NSF project OIA-1655280 and DOE project DE-SC0021391. Nathan Humphrey thanks Dr. Robert Arslanbekov for introducing him to plasma modeling and to simulations in COMSOL. The authors thank Dr. Evgeny Bogdanov for sharing his experiences with plasma simulations using COMSOL.